\documentclass{cernrep}
\usepackage{xcolor}
\usepackage{appendix}

\begin{document}

\vspace*{5 cm}

\begin{center}

{\huge \bf The Belle~II Experiment at SuperKEKB}

{\huge \bf Input to the European Particle Physics Strategy}

\vspace*{0.5 cm}

{\Large The Belle~II Collaboration}

\vspace*{0.3 cm}

\texttt{esppu-conveners@belle2.org}

\end{center}

\vspace*{0.5 cm}

Belle II is an intensity-frontier experiment at the SuperKEKB collider in Tsukuba, Japan. Over the coming decades, it will record the decays of billions of bottom mesons, charm hadrons, and tau leptons produced in 10 GeV electron-positron collisions. The experiment's low-background environment and precisely known kinematics enable high-precision measurements of hundreds of Standard Model (SM) parameters while probing for new particles at mass scales far beyond the direct reach of high-energy colliders. We project Belle II's sensitivity for key measurements - where it will be uniquely positioned or world-leading - over datasets ranging from 1 to 50 ab${}^{-1}$. By exploring previously uncharted regions of non-SM parameter space with high precision, Belle II will either reveal new physics or set stringent constraints, guiding future experimental and theoretical efforts. Additionally, we outline near-term upgrades to the Belle II detector and SuperKEKB accelerator, which will enhance sensitivity in searches for new physics beyond the SM across flavor, tau, electroweak, and dark sector physics. These improvements will ensure that Belle II remains both complementary to and competitive with the LHC and other experiments.

\newpage

\title{The Belle~II Experiment at SuperKEKB\newline
Input to the European Particle Physics Strategy 2025}
\author{The Belle~II Collaboration\thanks{Belle~II ESPP Preparatory Group: F.~Bernlochner, V.~Bertacchi, T.~E.~Browder, A.~Gaz, T.~Kuhr, P.~Lewis, J.~Libby, C.~Marinas, M.~Masuzawa, M.~Roney, K.~Trabelsi}}

\maketitle

\label{Sec1}

Belle~II is a current-generation flavor physics experiment that began collecting electron-positron collision data in 2018 (``Run 1'') at the SuperKEKB collider, the flagship facility of KEK in Tsukuba, Japan. With peak luminosities now exceeding $5.0 \times 10^{34} \,\text{cm}^{-2} \text{s}^{-1}$ and strong European participation, Belle~II is dedicated to high-precision measurements and searches for New Physics (NP), leveraging the unique advantages of its $e^+ e^-$ collision environment.

Data-taking resumed in 2024 (``Run 2'') and will continue until 2032, when a total integrated luminosity of 10~ab$^{-1}$ is expected. 
An upgrade of the detector is then planned to allow operation at higher instantaneous luminosity, targeting a total sample of 50~ab$^{-1}$. This sample will contain billions of decays of bottom mesons, charm hadrons, and tau leptons, enabling precise tests of the Standard Model (SM) and sensitive NP searches across a vast parameter space. Belle~II will measure rare decays, quark-mixing parameters, and search for dark-sector particles, potentially revealing new physics or setting stringent constraints on beyond-SM (BSM) theories.

Beyond its physics program, Belle~II prioritizes detector R\&D for an upcoming upgrade. These improvements will enhance its BSM physics sensitivity, while maintaining competitiveness with the LHC and other experiments. Major European institutions play a key role in this effort, which will also inform future collider experiment designs. The high luminosity and beam backgrounds at Belle~II provide an ideal testing ground for next-generation detector technologies, including those critical for future $e^+e^-$ colliders. A SuperKEKB upgrade is in development, reinforcing its role as a benchmark for future electron-positron colliders and contributing to $e^+ e^-$ Higgs factory initiatives.

With this letter, we emphasize the importance of continued precision SM studies and rare decay searches at ultra-high luminosities, particularly in the intermediate energy regime of $e^+e^-$ collisions, given their unique discovery potential. Furthermore, sustaining robust international collaborations — particularly beyond Europe, strengthens global scientific progress. In line with European Strategy recommendations, the U.S. Particle Physics Project Prioritization Panel (P5), and Japan’s MEXT Roadmap, we strongly advocate for the inclusion of Belle~II’s physics program and upgrade plans in the ESPPU as a vital component of the European flavor physics agenda.

Belle~II is an international experiment with about 700 collaborators from 124 institutions across 28 countries and regions, with Early Career Researchers comprising over 50\% of its membership. European institutions play a crucial role, with 34 universities and major national labs actively contributing to detector development, operations, upgrades, software, computing, physics, and performance.

The structure of the document is as follows. Section 1 summarizes our physics reach for three selected luminosity benchmarks. Section 2 contains a brief description of the SuperKEKB accelerator and the Belle~II detector, followed by a summary of current performance and future plans in Section 3. 

\newpage

\section{Physics objectives}
\label{Sec2}

The physics program of Belle~II is vast and broad \cite{Belle-II:2022cgf}, covering such disparate topics as CP violation, lepton universality, dark matter, new physics in rare decays, precision QCD, and quantum observables\footnote{``Quantum Information meets High-Energy Physics: Input to the update of the European Strategy for Particle Physics''}. For the purposes of this strategy document, we highlight three measurements that are feasible at each of three integrated luminosity benchmarks:

\begin{itemize}
\item $365$~fb$^{-1}$ (Run 1, completed),
\item 10~ab$^{-1}$,
\item 50~ab$^{-1}$.
\end{itemize}

These measurements were chosen to highlight Belle~II's unique capabilities relative to other experiments and to illustrate the evolving nature of its most impactful measurements in future data samples. In addition to these, a large number of additional precision parameter measurements, anomaly searches, and rare decay measurements will be unlocked or improved with increasing integrated luminosity. At the same time,  improvements in analysis tools, calibrations, and performance will continue to increase the physics impact of each unit of collected data. 

\subsection{Highlights from Run 1}
Of the 66 Belle~II publications so far from Run 1\cite{pubs}, we highlight the following as representative of the impact, uniqueness, and range of the Belle~II physics program. 

\subsubsection{Evidence for $\boldsymbol{B^+\to K^+\nu\bar{\nu}}$ (a new FCNC process)}  
Flavor-changing neutral current processes (FCNCs) such as $B^{+}\to K^{+}\nu\bar{\nu}$ only occur at higher orders in the standard model (SM). Thus, they are probes of possible virtual (beyond the SM) BSM contributions. Using the Run 1 data set, a machine learning-based analysis provided the first evidence for this decay \cite{Belle-II:2023esi}. We exploited the constrained kinematics of the initial state, the significant missing four-momenta of the neutrinos produced by the signal decay, and the low multiplicity in $e^+e^-$ collisions to develop an inclusive reconstruction with high efficiency. The inclusive analysis was validated by a complementary analysis that fully reconstructed the $B$ meson accompanying the signal candidate, which had lower efficiency but higher purity. We measure a branching fraction of
$(2.3\pm 0.5 \pm 0.5)\times 10^{-5}$. Here and elsewhere, the first uncertainty is statistical, and the second systematic. This is the first evidence of a $b\to s \nu \bar{\nu}$ decay, with a significance of 3.5 standard deviations. The result is 2.7 standard deviations above the SM prediction. 

This measurement of a $b\to s \nu \bar{\nu}$ process is not possible in proton-proton collisions. Further, it probes the possibility of BSM physics in the third generation of quarks and leptons. Systematic uncertainties will improve with larger data samples because they are mainly derived from data control channels.

\subsubsection{Precise measurement of the $\tau$-lepton mass ($\tau$ physics and lepton universality checks)}
Electron-positron collisions at a c.m.\! energy of the $\Upsilon(4S)$ resonance not only produce large samples of $B$-meson pairs; in addition, they also produce a nearly equivalent number of $\tau$-lepton pairs. Using a sample corresponding to 190~fb$^{-1}$, Belle II has reported the most precise measurement of the $\tau$-lepton mass $m_{\tau}$ to date \cite{Belle-II:2023izd}, which is one of the critical ingredients for a lepton universality test.

A pseudo-mass $M_{\min}$ distribution from selected $\tau^-\to \pi^-\pi^+\pi^-\nu$ candidates determines the $\tau$-lepton mass:
$$
 M_{\min}=\sqrt{M_{3\pi}^2+2\left(\tfrac{\sqrt{s}}{2}-E^{\ast}_{3\pi}\right)\left(E^{\ast}_{3\pi}-\left|\vec{\mathbf{p}}^{\ast}_{3\pi}\right|\right)}\leq m_{\tau}\;,
$$
where $M_{3\pi}$, $E^{*}_{3\pi}$, and $\mathbf{p}^{\ast}_{3\pi}$ are the invariant mass, energy, and momentum of the reconstructed $\pi^-\pi^+\pi^{+}$ system; the momentum and energy are calculated in the $e^+e^-$ c.m.\! frame. Here $\sqrt{s}$ is the $e^+e^-$ c.m.\! energy. The distribution ends at the $\tau$-lepton  mass, so Belle II used the position of this cutoff to determine the mass. An empirical function is fit to the distribution in the region of the cutoff, which accounts for resolution and the effects of initial- and final-state radiation. The position of the cutoff gives
$$
 m_{\tau}= 1777.09\pm 0.08 \pm 0.11~\mathrm{MeV}/c^2 \; .
$$
The systematic uncertainty is dominated by our knowledge of the $\sqrt{s}$ and $\left|\vec{\mathbf{p}}^{\ast}_{3\pi}\right|$ scales, which are calibrated using the known $B$- and $D$-meson masses, respectively.

Precision can be improved with more accurate c.m.\! energy- and momentum-scale calibrations, which will be possible with large samples and better knowledge of the $\Upsilon(4S)$ lineshape. However, current precision indicates a good systematic understanding of the Belle~II data, which has led to a three-fold improvement in precision compared to the Belle result \cite{Belle:2006qqw}.
\subsubsection{Search for a dark photon (dark sector physics)}
A dark photon $A^{\prime}$ is a postulated vector gauge boson from an extra dark $U(1)$ symmetry that couples to SM particles via kinetic mixing with the SM photon; the strength of the mixing is parameterized by a dimensionless constant $\epsilon$. The dark photon can couple to dark sector particles with dark U(1) charge. If the lightest dark sector particle $\chi$ is light enough so that $2m_{\chi} < m_{A^{\prime}}$, the dark photon will decay predominantly into a $\chi \chi^{\prime}$ pair. 

Belle~II has unique reach for the dark photon through its invisible signature: a single initial-state radiation photon with large missing energy. With Run 1 data, sensitivity to $\epsilon$ is expected in the range $10^{-3}$ to $10^{-4}$, which will significantly improve existing constraints \cite{Belle-II:2022cgf}. Sensitivity to dark photons with masses below 2~GeV$/c^2$ will depend on stringent vetoes of cosmic rays and QED backgrounds using the electromagnetic calorimeter and the muon and $K^{0}_{\rm L}$ detector; these veto studies are in progress. 

The model-independent limits on a single initial-state photon accompanied by missing energy will constrain regions of the parameter space for several dark-sector models consistent with the observed relic density \cite{Belle-II:2022cgf}. 
This measurement already shows the capability of Belle~II to contribute significantly to areas beyond flavor physics.

\subsection{Highlights with $\boldsymbol{10~\mathrm{ab}^{-1}}$ of data}
With 10~$\mathrm{ab}^{-1}$, new $B$-decay modes will be observed and several indirect searches for physics-beyond-the-SM will have discovery potential. We highlight the following three likely results that illustrate the impact of the Belle~II physics program at this benchmark.\footnote{In version 2 of this submission, we have added App.~\ref{app:tau} that describes our prospects for precision measurements of the tau lepton. These will be systematically limited with a $10~\mathrm{ab}^{-1}$ sample.}

\subsubsection{Discovery of $\boldsymbol{B^{+}\to\mu^+\nu}$ (virtual W-annihilation decay)}
Purely leptonic decays $B\to \ell \nu_\ell$, $\ell=e,\mu,\tau$ are suppressed by the CKM matrix-element $|V_{ub}|$ and have a helicity factor proportional to the squared lepton mass. The decay constant $f_{B}$ can be computed with a theoretical uncertainty of 0.7\% using lattice QCD~\cite{FlavourLatticeAveragingGroupFLAG:2024oxs}, making these modes sensitive probes for SM extensions involving an extended Higgs sector or leptoquarks, which would alter the decay rate. In addition to probing charged non-SM fields, these decays will enable complementary determinations of the CKM matrix element $|V_{ub}|$ that could be instrumental in resolving the discrepancy observed between measurements made with exclusive and inclusive semileptonic decays \cite{HeavyFlavorAveragingGroupHFLAV:2024ctg}. The predicted branching fractions of $B\to \tau \nu_\tau$, $B\to \mu \nu_\mu$, and $B\to e \nu_e$ are around $1 \times 10^{-4}$, $4 \times 10^{-7}$, and $9 \times10^{-12}$ in the SM, respectively; these predictions are based on global values of $|V_{ub}|$~\cite{HeavyFlavorAveragingGroupHFLAV:2024ctg} and $f_{B} = 190 \pm 1.3~\mathrm{MeV}/c^{2}$~\cite{FlavourLatticeAveragingGroupFLAG:2024oxs}.

Analyses are challenging, as these annihilation decays are rare and involve final states with missing energy from neutrinos and few tracks. However, the two-body $B^+\to \mu^+ \nu$ decay leads to a monochromatic muon in the rest frame of the signal $B$ meson, which is a distinctive experimental signature. This implies that the statistical precision in the $B^+\to \mu^+ \nu$ mode is competitive with $B\to \tau^+\nu$ despite the stronger helicity suppression. Precision studies of these modes is unique to Belle~II given their missing-energy signatures.  

Baseline projections based on Belle results using hadronic, semileptonic, and inclusive tagging, combined with demonstrated Belle~II improvements, suggest that Belle~II will observe $B\to \mu \nu_\mu$ decays with a data sample approximately equivalent to $10~\mathrm{ab}^{-1}$ \cite{Belle-II:2022cgf}, which will correspond to a relative $|V_{ub}|$ precision of around 5\%. 

\subsubsection{Measurements of branching fraction ratios $\boldsymbol{R(D^{\ast})}$ and $\boldsymbol{R(D)}$ (lepton flavor universality in $b\to c \tau \nu$ processes)} 
The decays $B \rightarrow D^{(*)} \tau \nu_{\tau}$ test lepton-flavor universality in the third generation, which can be violated if lower-mass (TeV range) BSM particles contribute to the process. Sensitive observables are the ratios $R(D)$ and $R(D^*)$ between branching fractions of $B \rightarrow D^{(*)} \tau \nu_{\tau}$ and $B\rightarrow D^{(*)} \ell \nu_{\ell}$ decays, where  $\ell=e$ or $\mu$. 
Current best results on $R(D^{(*)})$ are reported by the Belle experiment~\cite{Belle:2019rba}; the result of a global average with other measurements leads to a 3.3 standard deviation from the SM expectation~\cite{HeavyFlavorAveragingGroupHFLAV:2024ctg}.

Investigating the anomaly through precision measurements of $R(D^{(*)})$ is a chief goal of Belle~II. We have already published our first measurement of $R(D^{\ast})$ using a 189~fb$^{-1}$ data set \cite{Belle-II:2024ami}; the measurement has better per fb$^{-1}$ precision than previous measurements due to improved hadronic $B$ tagging and optimized selections. 

With a 10~ab$^{-1}$ data sample, several systematic challenges must be overcome. The most significant is a detailed understanding of poorly known $B\rightarrow D^{**} \ell \nu$ backgrounds, whose feed-down may bias the results. Our Run 2 data will allow for accurate tagged measurements of $B\rightarrow D^{**} \ell \nu$ decays for several $D^{**}$ states using samples that reconstruct a lepton, a $D^{(*)}$ meson and one or more pions on the signal side. 

Measurements with 10~ab$^{-1}$ will exclude the SM if the current central value persists. The relative precisions of these measurements will be approximately 1.8\% and 3\% for $R(D^{\ast})$ and $R(D)$, respectively. LHCb will also make measurements and anticipates a 5\% precision on $R(D^{\ast})$ before their Upgrade II \cite{LHCb:2021glh}. This is an area of competition between the experiments. However, we have an advantage due to our full-event reconstruction and our ability to measure $D^{\ast\ast}$ contributions.

 \subsubsection{CP~violation in $\boldsymbol{B^0\to K^0\eta^{\prime}}$ decays ($b\to s$ penguin processes)}
The decay $B^0 \to \eta'K_S^0$ has a sizable decay rate dominated by a $b\to s$ loop amplitude, where non-SM physics can contribute.  The quantity of interest is $\Delta{\cal S}_{\eta'K_S^0}\equiv{\cal S}_{\eta'K_S^0} -\sin 2\phi_1$, where ${\cal S}_{\eta'K_S^0}$ is the time-dependent $CP$ asymmetry in $B^0 \to \eta'K_S^0$ decays. Here, the CKM unitarity triangle angle $\phi_1$ (or $\beta$) is accurately predicted by CKM unitarity and measured in tree-amplitude-dominated decays.
SM predictions that include a systematic treatment of low-energy QCD amplitudes and assume factorization yield $0.00<\Delta{\cal S}_{\eta'K_S^0}<0.03$~\cite{Beneke:2005pu}. Establishing $\Delta{\cal S}_{\eta'K_S^0}$ is inconsistent with this prediction would indicate non-SM dynamics. The current global value of $\Delta {\cal S}_{\eta'K_S^0}^{\rm exp}$ is $-0.07\pm0.05$~\cite{HeavyFlavorAveragingGroupHFLAV:2024ctg}, including the first measurement from Belle~II using the Run I data set \cite{Belle-II:2024xzm}.  Manageable backgrounds and its high-resolution electromagnetic crystal calorimeter offer Belle~II unique access to this measurement. The precision of the SM prediction, along with the excellent Belle~II experimental perspectives, make $B^0\to\eta'K_S^0$ the most promising channel in this program.

With a 10~ab$^{-1}$ sample, we anticipate an uncertainty in $\Delta {\cal S}_{\eta'K_S^0}^{\rm exp}$ of 0.02 or better \cite{Belle-II:2022cgf}, which is similar to the theoretical uncertainty. Thus, this measurement will be a stringent test of BSM contributions to $CP$~violation.

\subsection{Highlights with $\boldsymbol{50~\mathrm{ab}^{-1}}$ of data}
A $50~\mathrm{ab}^{-1}$ data sample is the ultimate goal for Belle~II, but will require an upgraded detector and SuperKEKB accelerator to achieve. We highlight three likely results that illustrate the impact of the Belle~II physics program at this benchmark. 

\subsubsection{CKM unitarity triangle angle $\boldsymbol{\phi_2}$ (also known as $\alpha$)}
Studies of charmless $B$ decays give access to $\phi_2$, the least known angle of the CKM unitarity triangle, and probe contributions of BSM dynamics in processes mediated by loop decay-amplitudes. To overcome hadronic uncertainties, combinations of measurements from decays related by isospin symmetry yield robust direct determinations of $\phi_2$~\cite{Charles:2017evz}.  Belle~II has the unique capability of studying jointly, and within the same experimental environment, all relevant final states of isospin-related charmless decays.

The most promising determination of $\phi_2$ relies on the combined analysis of the decays $B^+ \to  \rho^+\rho^0$, $B^0 \to\rho^+\rho^-$, $B^0\to \rho^0\rho^0$, and the corresponding decays into pions. The current global precision of 4 degrees is dominated by $B \to \rho\rho$ data~\cite{HeavyFlavorAveragingGroupHFLAV:2024ctg}.  Efficient reconstruction of low-energy $\pi^0$ mesons makes improved measurements in $B^+\to \rho^+\rho^0$ and $B^0 \to \rho^+\rho^-$ decays (which has two $\pi^0$'s in the final state) possible only at Belle~II. 

The expected Belle~II performance has been demonstrated by a recent $B^0\to \rho^+\rho^-$ analysis on par with the world's best results~\cite{Belle-II:2024frs}. Contributions from model assumptions on poorly known decays involving $a_1$, $f_0$, and non-resonant $\pi\pi$ final states will be mitigated by amplitude analyses.  

Complementary $\phi_2$ determinations based on $B \to \pi\pi$ decays are limited by poor knowledge of the branching fraction and direct $CP\!$-violating asymmetry of the $B^0 \to \pi^0\pi^0$ decay. Belle~II is the only experiment with good sensitivity for this all-neutral channel. Current Belle~II results~\cite{Belle-II:2024baw} are already comparable to the world's best results. Since the only irreducible systematic uncertainty is associated with the (sub-percent) precision of branching fractions of control channels used to determine the $\pi^0$ reconstruction efficiency from data, this measurement has an approximate ten-fold margin of improvement when extrapolated to the full Belle~II data set.

An auxiliary measurement of the decay-time-dependent $CP\!$-violating asymmetry in $B^0 \to \pi^0\pi^0$ decays improves $\phi_2$ knowledge by resolving mirror solutions. With the full Belle~II sample, we can carry out time-dependent analysis with photon conversions and $\pi^0$ Dalitz decays~\cite{Belle-II:2018jsg}. Combining all the information from $B\to\rho\rho$ and $B\to\pi\pi$ decays, the ultimate uncertainty on $\phi_2$ would be 0.6~degrees with 50\,ab$^{-1}$ of integrated luminosity \cite{Belle-II:2022cgf}. This sensitivity may improve further with the inclusion of $B\to \rho\pi$ modes, which also resolve ambiguities in $\phi_2$ solutions\cite{Belle-II:2018jsg}.

\subsubsection{Tests of $\boldsymbol{CP}$\! violation in charm-meson decay (search for BSM in charm)}
The first observations of $CP\!$ violation in charm decay was a $CP\!$-asymmetry difference between $D^0\to K^+ K^-$ and $D^0\to\pi^+\pi^-$ decays~\cite{LHCb:2019hro}; this has been shown to be due to a significant asymmetry in $D^0\to \pi^+ \pi^-$ decays \cite{LHCb:2022lry}. In contrast to $B$ decays, loop amplitudes in charm are severely suppressed by the Glashow-Iliopoulos-Maiani (GIM) mechanism. Standard model $CP\!$ violation arises mostly from the interference of tree-level amplitudes, possibly associated with rescattering~\cite{Grossman:2019,Bediaga:2022sxw}. Rescattering amplitudes are challenging to compute and make the interpretation of the observed $CP\!$ violation ambiguous~\cite{Chala:2019}. 

Precise measurements of $CP\!$ asymmetries in other decay channels are crucial to understand the underlying dynamics. Cabibbo-suppressed decays such as $D^+ \to \pi^+\pi^0$ and $D^0 \to \pi^0\pi^0$ are particularly interesting due to their different isospin transitions compared to $D^0 \to \pi^+\pi^-$. 
 In particular, a probe for BSM physics is provided by an isospin-based sum rule that relates branching fractions, $CP\!$ asymmetries, and total widths of $D^0\to\pi^+\pi^-$, $D^+\to\pi^+\pi^0$, and $D^0\to\pi^0\pi^0$ decays to discriminate whether observations of $CP\!$ violation in the individual channels are due to BSM physics or not~\cite{Grossman:2012eb}. 
 The properties of the $D^0\to\pi^0\pi^0$ decay at the required level of precision will only be accessible at Belle~II.

The expected sensitivities to charm $CP$ violation using $D^{*+}$-tagged decays reconstructed in the currently available Belle~II data, extrapolated to a 50~ab$^{-1}$ sample are 0.13\% for $D^{+}\to \pi^+\pi^0$ \cite{pipi0inpreparation} and 0.07\% for $D^0\to \pi^0\pi^0$ \cite{Belle-II:2022cgf}. Only the statistical sensitivity is considered, as systematic uncertainties will be small. Additional sensitivity will come from untagged $D^+$ decays and from $D^0\to\pi^0\pi^0$ decays where the $D^0$ flavor is inferred from the rest of the event or from flavor-specific $B \to D X$ decays~\cite{Belle-II:2023vra}. 

\subsubsection{Search for lepton-flavor violation in $\boldsymbol{\tau^-\to\ell^-\gamma}$ (rare $\tau$ lepton decays)}
No charged lepton flavor violation has been observed.  Minimal SM extensions that include right-handed neutrinos enable lepton-flavor violation, but through heavily suppressed mechanisms,  which would yield branching ratios too small to be observed in current and foreseen experiments. Extensions with BSM interactions
predict lepton-flavor-violating $\tau$ decays at the $10^{-10}$--$10^{-8}$ levels, which will be probed at Belle~II with the approximately 50 billion tau pairs produced in a $50~\mathrm{ab}^{-1}$ sample.

The decays $\tau^-\to e^-\gamma$ and $\tau^-\to\mu^- \gamma$ are considered particularly promising as they are predicted to occur at rates very close to the current experimental limits in a wide class of non-SM theories. We consider the presence of irreducible backgrounds from leptonic tau decay and QED backgrounds for $\tau^- \to \ell^-\gamma$ decays, thus the precision is proportional to ${\cal{L}}^{-1/2}$. The expected sensitivity is $(7-8)\times 10^{-9}$ \cite{Belle-II:2022cgf}. This precision may be improved by selections using machine-learning (ML) techniques, which have recently been demonstrated with our world leading limits on $\tau^{-}\to \mu^{-}\mu^{+}\mu^{-}$ decays \cite{Belle-II:2024sce}.

\section{SuperKEKB and Belle~II}
\label{Sec3}

The SuperKEKB accelerator \cite{Onishi} is designed to collide electrons and positrons at center-of-mass energies in the region of the $\Upsilon$ resonances. The beam energies for the High Energy Ring (HER) and the Low Energy Ring (LER) are 7 GeV and 4 GeV, respectively, with a target luminosity of $6.0 \times 10^{35} \text{ cm}^{-2} \text{s}^{-1}$. SuperKEKB is the world's most powerful electron-positron collider, leading the exploration of the luminosity frontier.

Beam collisions with a vertical beam spot size of $\mathcal{O}(100)$ nanometers at the interaction point produce billions of bottom, charm, and tau pairs within the Belle~II detector. The accelerator’s asymmetric beam energies provide a boost of $\beta\gamma = 0.28$ of the center-of-mass system, enabling time-dependent CP violation measurements. Most data will be collected at the $\Upsilon(4S)$ resonance ($\sqrt{s} = 10.58$ GeV), just above the threshold for $B$-meson pair production. In $B$-meson pair production, no additional fragmentation particles are generated, and since the collisions occur in a well-defined initial state with known four-momentum, Belle~II excels in analyses involving neutral particles and inclusive final-state reconstruction.

The Belle~II detector, described in \cite{tdr}, is a near-hermetic magnetic spectrometer comprising several subdetectors surrounding the interaction region:

\begin{itemize}
    \item \textbf{Vertex Detector (VXD):} Consists of two subdetectors: a Pixel Vertex Detector (PXD) with two layers of DEPFET pixel sensors and a Silicon Vertex Detector (SVD) with four layers of double-sided silicon strip sensors. The VXD is used primarily to find the vertices of charged tracks.
    \item \textbf{Central Drift Chamber (CDC):} A large-volume gas-based tracking system surrounding the VXD, providing precise momentum measurements.
    \item \textbf{Particle Identification System:} Consists of two Cherenkov detectors: the Time-Of-Propagation (TOP) system in the barrel region, which measures the propagation time and impact positions of Cherenkov photons, and the Aerogel Ring Imaging Cherenkov (ARICH) detector in the forward region, which detects the number and positions of Cherenkov photons.
    \item \textbf{Electromagnetic Calorimeter (ECL):} Constructed from CsI(Tl) crystals, it detects and measures the energies of photons and identifies electrons. The calorimeter electronics have been optimized to reduce pile-up noise.
    \item \textbf{K-Long and Muon (KLM) Detector:} Located outside the 1.5 T superconducting solenoid, it includes resistive plate chambers and scintillator strips with silicon photomultipliers that operate efficiently under high neutron fluxes. The KLM identifies muon tracks and clusters from K-Long showers. 
\end{itemize}

The trigger and data acquisition (DAQ) systems are configured to handle high event rates and support Belle~II’s extensive physics program. Data processing utilizes a grid-based computing system~\cite{grid}.

\section{Status and outlook}
\label{Sec4}

The commissioning of the accelerator and detector started in February 2018. Currently, the accelerator is operating with the superconducting final focus. SuperKEKB reached a peak luminosity of 5.1$\times$10$^{34}$ cm$^{-2}$s$^{-1}$ with stored currents of 1354 mA in HER and 1699 mA in LER. The novel nano-beam scheme, which is based on a large horizontal crossing angle, was successfully implemented with high specific luminosity even when $\beta^*_y < \sigma_z$, where $\sigma_z$ corresponds to the longitudinal bunch length. SuperKEKB uses a special "crab waist" scheme to avoid beam-beam instabilities. All of the data taking periods have included an operational Belle~II detector, collecting data with an overall efficiency reaching 90 $\%$. To date, the Belle II experiment was able to integrate about $575$~fb$^{-1}$. Collimator positions have been optimized and background levels are moderate. 
 
As discussed further below, these initial running periods have been challenging, and significant run time has been devoted to luminosity and background optimization. Belle II/SuperKEKB plans to run for 7 months/year with a 4 month summer shutdown to avoid high electricity costs. SuperKEKB plans to reach a peak luminosity of 1.0$\times$10$^{35}$ cm$^{-2}$s$^{-1}$ by 2026 and, with further upgrades, achieve an integrated luminosity of $50$~ab$^{-1}$.

In terms of Belle~II operation and performance, the observed impact-parameter resolution is 10–15 µm, resulting in 20–30 µm typical vertex resolution. The relative charged-particle transverse momentum resolution is approximately 0.3\% at $p_T = 1$~GeV/c. The observed hadron identification efficiencies are 90$\%$ at 10$\%$ contamination. Uncertainties in hadron-identification performance are of the order of 1$\%$. The energies of electrons and photons are measured with energy-dependent resolutions in the 1.6–4$\%$ range. Our observed lepton-identification performance is 0.5$\%$ pion contamination at 90$\%$ electron efficiency, and 7$\%$ kaon contamination at 90$\%$ muon efficiency. Typical uncertainties in lepton-identification performance are at the 1$\%$–2$\%$ level.

\vspace{0.5cm}

\subsection{Challenges for full luminosity running}

The most recent run of the SuperKEKB accelerator (called 2024c, from October until December 2024) focused on increasing the peak luminosity and reducing unexpected sudden beam loss (SBL). The operation of the machine with the novel nano-beam scheme has been challenging. Much time and effort was spent on machine tuning to overcome several of the observed issues: beam emittance blowup, beam-beam effects and SBLs. There were also frequent hardware problems with the RF gun used for HER electron injection. 
However, continuous improvements led to a new luminosity world record and confidence in a plan to reach even higher luminosities. 

During this operation period, due to difficulties with injection, the vertical beta function $\beta_y^*$ at the interaction point was chosen to be 1~mm.
Beam-beam effects caused a deterioration in the LER injection efficiency. However, by
squeezing the LER horizontal beta function at the interaction point (from 80 mm down to 60 mm) at the end of the run, it became possible to increase the stored beam currents. The possibility of increasing the stored beam currents together with more refined optics corrections led to the final peak luminosity figures.

The problem of SBLs in the LER accompanied by pressure bursts in the wiggler section 
was a limiting factor for much of recent SuperKEKB operation. The frequency of SBLs was greatly reduced by cleaning the inside of the beam pipe flanges in a key wiggler section, but SBLs accompanied by quenches of the superconducting magnet for final focusing (QCS) still occurred. Fortunately, there was no serious damage to accelerator components such as collimators, and we were able to complete the 2024c operation as scheduled.

The main detector challenges for full luminosity run are expected to be related to the level of the beam background from several sources:  beam-gas interactions, Touschek intra-beam scattering, radiative Bhabha events, low energy electron-positron pairs from two-photon events, and injection backgrounds. The background levels measured during the past runs can be used to understand and estimate the harsh conditions that Belle~II will face, but since the machine configuration and optics will further evolve, background extrapolations at final luminosity still have large uncertainties.

\vspace{0.5cm}

\subsection{Future opportunities and R\&D programs}

Improvements in instantaneous luminosity will accelerate data accumulation, the Belle~II experiment is expected to operate for many years. The Belle~II and SuperKEKB communities are already exploring future accelerator and detector upgrades to maximize physics sensitivity at the highest luminosities and background environments. Background mitigation remains the most critical challenge for long-term operation, as several subdetectors are sensitive to increased radiation and background levels. To address this, planned upgrades \cite{{cdr}} focus on reducing occupancy levels through higher granularity and faster timing, as well as improving radiation tolerance and robustness against single event upsets. These efforts are being developed within the DRD Collaborations, in which Belle~II groups are actively involved.

Ongoing upgrade development for each Belle~II subdetector is detailed in \cite{cdr}. The innermost detector systems, the PXD and SVD, will be fully replaced after Run 2 with a single-technology, depleted monolithic CMOS active pixel sensors. Compared to the current PXD, this new vertex system will offer superior spatial and time resolution, and enhanced radiation hardness. Given the high background levels, an extension of the silicon sensors is under study to replace the inner layers of the CDC. For the outer CDC region, additional R\&D efforts aim to mitigate the impact of increased backgrounds. In the TOP detector, a staged replacement of the MCP-PMTs with lifetime-extended sensors is planned to minimize quantum efficiency degradation from accumulated radiation dose. The ECL will retain its crystals and photosensors, but a new readout board is being developed to minimize pileup noise and improve trigger granularity. For the KLM, efforts are underway to mitigate efficiency degradation from higher background rates, including the development of a new, environmentally friendly gas mixture and upgraded readout electronics for the resistive plate chambers.

Belle II's software framework for simulation, reconstruction, and analysis, \texttt{basf2} \cite{{Kuhr:2018lps}, {basf2}}, will continue to be used. Algorithmic improvements and new computing architectures will be developed to adapt to increased data rates, occupancy, and background levels. 

Belle II will continue to leverage recent advances in AI/ML to improve the sensitivity of physics analyses and the performance of detector reconstruction. In addition, new ML on the front end will be used in the trigger. AI/ML techniques will replace some of the work by operators in tuning SuperKEKB. Some recent highlights in this domain include GNN (Graph Neural Network)-based tracking~\cite{GNN_tracking} and clustering~\cite{GNN_clustering} algorithms, a FPGA-based neural network for single-track triggers~\cite{NN_trigger}, and ML-based tuning of the linac. 

On the machine side, the next operation phase of the SuperKEKB main ring is scheduled to begin in fall 2025, with several upgrades already underway. Cleaning of the LER beam pipe's inner surface in all the sections is planned as an SBL countermeasure, and reinforcement of radiation shielding in the Oho straight section will enable full-scale use of the nonlinear collimator. A new RF gun for the HER is being prepared. 

A detailed plan has been developed to reach a peak luminosity of $1 \times 10^{35} \, \text{cm}^{-2} \text{s}^{-1}$ and an integrated luminosity of 1 ab$^{-1}$ by implementing optics squeezing ($\beta^*_y = 0.9$ mm) and increasing stored currents (2.6 A for LER and 1.8 A for HER). Before achieving this goal, injection efficiency must be improved and stabilized against beam-beam effects. Further reduction of $\beta^*_y$ to 0.6 mm, while increasing LER/HER beam currents to 3.0 A and 2.0 A, respectively, is planned before LS2. To accomplish this, optimizing rotatable sextupole magnet configurations is essential for correcting chromatic x-y coupling and improving the dynamic aperture. It is also important to reduce couplings and aberrations at the IP. One of the key challenges is understanding why the specific luminosity is lower than the predictions of simulation, particularly at high bunch currents. Identifying and addressing this discrepancy is critical. 

While continuing accelerator operations, upgrade objectives will be evaluated for the next long shutdown (LS2), alongside necessary R\&D efforts. Upgrade efforts include improvements to the injector system, upgrades to the RF systems—such as increasing the number of klystrons and mitigating aging effects—and the interaction region (IR). The IR upgrades will require the development of superconducting final focusing magnets using Nb$_3$Sn cables, among other advancements. R\&D on the IR upgrade has started. The technical feasibility of these upgrades will be assessed over the next few years. It is important to recognize that the solutions developed to achieve design performance at SuperKEKB will likely have implications for the design of a future $e^+ e^-$ Higgs factory.

\vspace{0.5cm}

\subsection{Chiral Belle: Beam polarization upgrade}

Consideration is being given to a modest upgrade of SuperKEKB that enables the Belle~II detector to explore topics in precision electroweak physics with longitudinally polarized electrons. As described in a Snowmass White Paper~\cite{chiral}, Belle~II would measure five values of $\sin^2\theta_W$ via left-right asymmetry measurements ($A_{LR}$) in $e^+e^-\rightarrow e^+e^-$, $e^+e^-\rightarrow \mu^+\mu^-$,~ $e^+e^-\rightarrow \tau^+\tau^-$, ~$e^+e^-\rightarrow c\overline{c}$, and $e^+e^-\rightarrow b\overline{b}$ events. With 70\% polarization and 10's of ab$^{-1}$ of data,  $A_{LR}$ measurements would yield values of the neutral current (NC) coupling constant of each fermion species that match ($e$,$\tau$) or greatly exceed ($b$, $c$, $\mu$) the precision of existing $Z^0$ world averages, but at 10~GeV. This provides unique sensitivity to new physics revealed through deviations from the SM running of $\sin^2\theta_W$, such as would be expected from a dark sector analogue of the $Z^0$. The polarization upgrade also enables a set of other precision $\tau$ lepton measurements, including its anomalous magnetic moment form-factor, with unrivaled precision, thereby providing a measurement analogous to that of the $(g-2)$ of the muon, but in the third generation.

The upgrade involves installing a polarized source that injects transversely polarized electrons into the HER; spin rotator units on both sides of the interaction region (IR) to rotate the spin to the longitudinal direction and then return it to its stable transverse orientation; and a Compton polarimetry system. The spin rotator system replaces 4 existing normal-conducting arc dipoles (2 on either side of the IR) with 4 multi-function superconducting magnets. These new magnets have overlapping dipole and solenoid fields overlaid with six skew quadrupoles. Detailed multiple term tracking studies show that this solution yields polarization lifetimes that are significantly longer than the beam lifetime. R\&D on the magnets, Compton polarimeter and polarized source is currently underway. This program requires SuperKEKB to have both its high design luminosity as well as polarized beams. Once the R\&D is completed this decade, Chiral Belle could be deployed during the 2030s.

\vspace{0.5cm}






\newpage
\appendix
\section{Belle II projection for tau observables}\label{app:tau}
This appendix provides projections for tau physics observables related to the test of lepton flavour universality.
Contrary to the search for lepton flavour violating $\tau$ decays reported in the contribution~\cite{ATLAS:2025lrr}, the precision on these measurements is limited by systematic uncertainties. The extrapolation to higher future luminosities is thus based on some assumptions. 

\subsection{Mass and lifetime measurement}
Belle II have reported the most precise current measurement of the tau mass, which is $1777.09 \pm0.08 \pm 0.11$ MeV/c$^2$, using 190 fb $^{-1}$~\cite{Belle-II:2023izd}.
The main systematic uncertainties are coming from the beam energy and the momentum corrections. 
We assume that a factor-of-two improvement in systematic uncertainties will be achieved in the future compared to the values used in ~\cite{Belle-II:2023izd}. Ongoing studies confirm the feasibility of achieving this reduction in the near future.

The most precise determination of the tau lifetime has been obtained by Belle with 711fb $^{-1}$,  giving $290.17\pm 0.53\pm0.33$ fs~\cite{Belle:2013teo} using both $\tau$ decaying into 3 pions (so-called $3\times3$ topology).
The projections we present here are based on an updated analysis using Belle II simulation, which anticipates a significant reduction in systematic uncertainties compared to Belle, while also improving statistical precision by employing the more abundant $3\times 1$ topology. This is made possible thanks to the reduced beam spot size at the interaction point, which is unique to the nano-beam scheme of the SuperKEKb accelerator.
The largest contribution to the systematic error is expected to come from the statistics of the simulated sample, that will scale as $\sqrt\mathcal{L}$. We expect a further improvement on the other systematic uncertainties by a  factor two.

\subsection{Leptonic branching fraction}
The Belle II experiment has measured the ratio  of the  muonic and electronic tau decays with 362 $fb^{-1}$ to be $R_\mu = 0.9675 \pm 0.0007 \pm 0.0036$. In that case, the main source of uncertainty is coming from the efficiency of lepton identification. We also assume a factor two reduction of the systematic uncertainties for this analysis, that will come both from a better treatment of the correlations of lepton identification uncertainties, and the increase in statistics of the control samples used to determine the muon identification efficiency.

The absolute branching fractions $\mathcal{B}(\tau\to e\nu_e\nu_\tau)$ and $\mathcal{B}(\tau\to \mu\nu_\mu\nu_\tau)$ have not been measured at $B$-factories. 
Dedicated studies shows that Belle II could reach a statistical uncertainty of 0.008 \% for both modes, with a systematic uncertainty of 0.092 (0.095)\% for the electronic (muonic) final state with Run 1 data. We assume a factor two reduction of the systematic uncertainty for future measurements.

\subsection{Test of lepton flavour universality}
\begin{figure}[htb]
    \centering
   \includegraphics[width=0.85\columnwidth]{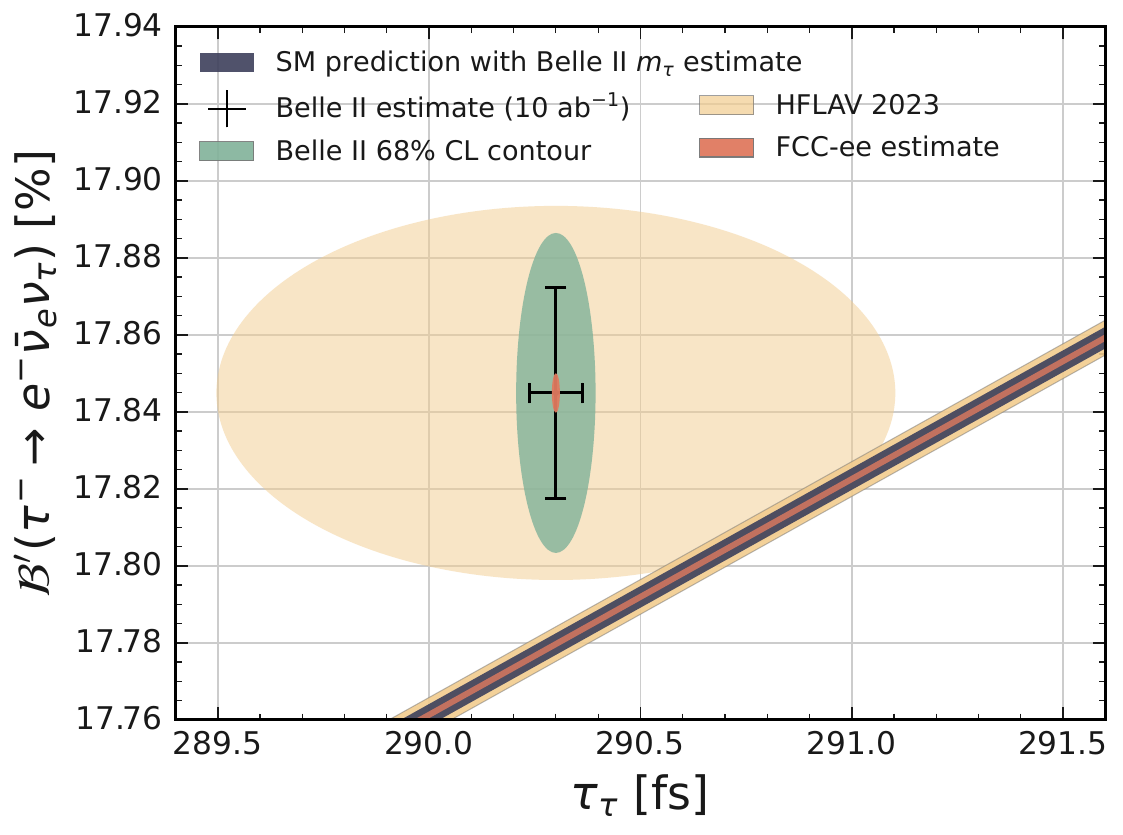}
    \caption{Comparison of the measured tau leptonic branching fraction and lifetime with the SM prediction, using Belle II expectations for 10 ab$^{-1}$, averaged with other measurements from the PDG~\cite{ParticleDataGroup:2024cfk}. Here, $\mathcal{B^ \prime}$ is the average of the direct $\mathcal{B}(\tau\to e\nu_e\nu_\tau)$ measurement with $\mathcal{B}(\tau\to \mu\nu_\mu\nu_\tau) $ assuming electron and muon have similar weak couplings. The latest HFLAV results~\cite{HeavyFlavorAveragingGroupHFLAV:2024ctg} as well as the projections of FCCee are also shown. }
    \label{fig:LFU}
\end{figure}

The expected precision for the tau lifetime, tau mass, the leptonic branching fractions and $R_\mu$ can be seen in Table~\ref{tab:projection} for 5 and 10 ab$^{-1}$. Given the systematic limitations on these measurements, additional statistics will bring little improvement on the results.
Those projections can be used to test lepton universality, as seen in Fig.~\ref{fig:LFU}, following the method from~\cite{HeavyFlavorAveragingGroupHFLAV:2024ctg}.

\begin{table}[h]
\caption{Projection of the total uncertainty on the observables for 5 and 10 ab$^{-1}$.}
\begin{center}
\begin{tabular}{c|ccccc}
Observables & lifetime (fs) & mass (MeV/c$^2$) & $\mathcal{B}(\tau\to e\nu_e\nu_\tau)$ (\%) & $\mathcal{B}(\tau\to \mu\nu_\mu\nu_\tau) (\%)$ &$R_\mu$ \\
\hline
5 ab$^{-1}$ &  0.064  & 0.064 & & & \\ 
 10 ab$^{-1}$ &   0.062 & 0.056 &  0.046 & 0.0475 & 0.0018 

\label{tab:projection}
\end{tabular}
\end{center}
\end{table}

\begin{thebibliography}{99}
\bibitem{Belle-II:2022cgf} L.~Aggarwal \textit{et al.} [Belle~II], [arXiv:2207.06307 [hep-ex]].
\bibitem{pubs} https://www.belle2.org/research/physics/publications
\bibitem{Belle-II:2023esi} I.~Adachi {\it et al.} [Belle~II], Phys. Rev. D 109, 11 (2024)
\bibitem{Belle-II:2023izd} I.~Adachi {\it et al.} [Belle~II], Phys. Rev. D 108, 3 (2023)
\bibitem{Belle:2006qqw} K.~Abe {\it et al.} [Belle~II], Phys. Rev. Lett. 99, 011801 (2007)
\bibitem{FlavourLatticeAveragingGroupFLAG:2024oxs} Y. Aoki {\it et al.}, Flavour Lattice Averaging Group (FLAG), arXiv:2411.04268 (2024)
\bibitem{HeavyFlavorAveragingGroupHFLAV:2024ctg} S.~Banerjee {\it et al.}, Heavy Flavor Averaging Group (HFLAV), arXiv: 2411.18639 (2024)
\bibitem{Belle:2019rba} G.~Caria {\it et al.} [Belle~II], Phys. Rev. Lett. 124, 16 (2020)
\bibitem{Belle-II:2024ami} I.~Adachi {\it et al.} [Belle~II], Phys. Rev. D 110, 7 (2024)
\bibitem{LHCb:2021glh} LHCb Collaboration, CERN-LHCC-2021-012, LHCB-TDR-023 (2021) 
\bibitem{Beneke:2005pu} M.~Beneke, Phys. Lett. B 620, 143 (2005)
\bibitem{Belle-II:2024xzm} I.~Adachi {\it et al.} [Belle~II], Phys. Rev. D 110, 11 (2024)
\bibitem{Charles:2017evz} J.~Charles, O.~Deschamps, S.~Descotes-Genon, and V.~Niess, Eur. Phys. J. C 77, 8 (2017)
\bibitem{Belle-II:2024frs} I.~Adachi {\it et al.} [Belle~II], arXiv:2412.19624 (2024)
\bibitem{Belle-II:2024baw} I.~Adachi {\it et al.} [Belle~II], arXiv:2412.14260 (2024)
\bibitem{Belle-II:2018jsg} E.~Kou {\it et al.}, PTEP 2019, 12 (2019)
\bibitem{LHCb:2019hro} R.~Aaij {\it et al.} (LHCb Coll.), Phys. Rev. Lett. 122, 21 (2019)
\bibitem{LHCb:2022lry} R.~Aaij {\it et al.} (LHCb Coll.), Phys. Rev. Lett. 131, 091802 (2023)
\bibitem{Grossman:2019} Y.~Grossman and S.~Schacht, J. High Energy Phys. 7, 20 (2019)
\bibitem{Bediaga:2022sxw} I.~Bediaga, T.~Frederico, and P.~Magalhaes, arXiv:2203.04056 (2022)
\bibitem{Chala:2019} M.~Chala,  A.~Lenz, A.~V.~Rusov, and J.~Scholtz, J. High Energy Phys. 7, 161 (2019)
\bibitem{Grossman:2012eb} Y.~Grossman, A.~L.~Kagan, and J.~Zupan, Phys. Rev. D 85, 114036 (2012)
\bibitem{pipi0inpreparation} Belle~II Coll., paper in preparation.
\bibitem{Belle-II:2023vra} I.~Adachi {\it et al.}, Phys. Rev. D 107, 11 (2023)
\bibitem{EWfit} J. Haller {\it et al.} (Gfitter Group), Eur. Phys. J. C78, 675 (2018). 
\bibitem{LHC750} ATLAS Coll., ATLAS-CONF-2015-081; CMS Coll., CMS-PAS-EXO-15-004.
\bibitem{CMSmumu} CMS Coll., CMS-HIG-16-017, CERN-EP-2018-204, arXiv:1808.01890 (2018). 
\bibitem{LFU} Several measurements of R(D(*)) by BaBar Coll., Belle Coll. and LHCb Coll. in the period 2012 - 2018, summarized by HFLAV, https://hflav.web.cern.ch/
\bibitem{B2TiP} E. Kou {\it et al.} [Belle~II], arXiv:1808.10567 (2018). 
\bibitem{Onishi} Y. Ohnishi et al., Progress of Theoretical and Experimental Physics, 2013 (3) (2013) 03A011
\bibitem{tdr} T. Abe \textit{et al.}, KEK-REPORT-2010-1, arXiv:1011.0352
\bibitem{Belle-II:2024sce} I.~Adachi {\it et al.} [Belle~II], JHEP 09, 062 (2024)
\bibitem{Kuhr:2018lps} T.~Kuhr \textit{et al.} [Belle~II Framework Software Group], Comput. Softw. Big Sci. \textbf{3} (2019) no.1, 1
\bibitem{chiral} A. Accardi \textit{et al.}, arXiv:2205.12847
\bibitem{cdr} H. Aihara \textit{et al.}, arXiv:2406.19421
\bibitem{basf2} The Belle II Collaboration, https://doi.org/10.5281/zenodo.14710811
\bibitem{GNN_tracking} L. Reuter \textit{et al.}, arXiv:2411.13596
\bibitem{GNN_clustering} F. Wemmer \textit{et al.}, Comp. Soft. Big Sci. 7, 13 (2023)
\bibitem{NN_trigger} S. B{\"a}hr {\textit et al.}, NIM-A 1073, 170279 (2025)
\bibitem{grid} V. Bansal, arXiv:1511.06760
\bibitem{ATLAS:2025lrr} [ATLAS, Belle II, CMS and LHCb colls.], arXiv:2503.24346.
\bibitem{Belle:2013teo} K.~Belous et al. [Belle coll.], Phys. Rev. Lett. {\bf 112}, 031801 (2014). 
\bibitem{ParticleDataGroup:2024cfk} S.~Navas et al. [Particle Data Group], Phys. Rev. D {\bf 110}, 030001 (2024).

\end{thebibliography}
\end{document}